\documentclass[aps,prl,reprint,groupedaddress]{revtex4-1}
\usepackage{graphicx}
\usepackage{amsmath}
\usepackage{amsfonts}
\usepackage{amssymb}
\usepackage[english]{babel}

\begin{document}


\title{Temporal-mode-selective optical Ramsey interferometry via cascaded frequency conversion}

\author{Dileep V. Reddy}
\email{dileep@uoregon.edu}
\author{Michael G. Raymer}
\affiliation{Oregon Center for Molecular, Optical and Quantum Science and Department of Physics, University of Oregon, Eugene, Oregon 97403, USA}

\maketitle \textbf{Temporal modes (TM) are a new basis for storage and
  retrieval of quantum information in states of light
  \cite{herit,Brecht:15prx}. The full TM manipulation toolkit requires
  a practical quantum pulse gate (QPG) \cite{eck11,silb11}, which is a
  device that unitarily maps any given TM component of the optical
  input field onto a different, easily separable subspace or degree of
  freedom \cite{Huang:13}. An ideal QPG must ``separate'' the selected
  TM component with unit efficiency, whilst avoiding crosstalk from
  orthogonal TMs \cite{Reddy:2013ip}. All attempts at implementing
  QPGs in pulsed-pump traveling-wave systems have been unable to
  satisfy both conditions simultaneously
  \cite{Zheng:01,Zheng:02,silb14,Ansari:16ax,Manurkar:16,Reddy:17,Huang:17}. This
  is due to a known selectivity limit in processes that rely on
  spatio-temporally local, nonlinear interactions between pulsed modes
  traveling at independent group velocities \cite{Reddy:2013ip}. This
  limit is a consequence of time ordering in the quantum dynamical
  evolution \cite{silb11,Quesada:pra14}, which is predicted to be
  overcome by coherently cascading multiple stages of low-efficiency,
  but highly TM-discriminatory QPGs
  \cite{Reddy:2014bt,Reddy:15pra,Quesada:16}. Multi-stage
  interferometric quantum frequency conversion in nonlinear waveguides
  was first proposed for precisely this purpose
  \cite{Reddy:2014bt}. TM-nonselective cascaded frequency conversion,
  also called optical Ramsey interferometry, has recently been
  demonstrated with continuous-wave (CW) fields
  \cite{Clemmen:16prl,Kobayashi:17}. Here, we present the first
  experimental demonstration of TM-selective optical Ramsey
  interferometry and show a significant enhancement in TM selectivity
  over single-stage schemes.}

  All-optical quantum frequency conversion (QFC) by three- or
  four-wave mixing in nonlinear materials is well known to preserve
  the quantum state of light \cite{huang92,mcg10a}. These processes
  are in principle noiseless, and can be used at sub-unity conversion
  efficiencies (CE) to generate single-photon color-superposition
  states across disjoint frequency bands. This facet has been posited
  as a two-level Hilbert space for single-photon qubits
  \cite{Clemmen:16prl}. Two QFC stages, each of which is set up for
  $50$\% CE, can be cascaded into a two-color interferometer
  constructed out of frequency-shifting beamsplitters. This effect has
  been shown for single-photon states in dual-pumped third-order
  nonlinear optical fibers \cite{Clemmen:16prl}, as well as for weak
  coherent states in singly-pumped second-order nonlinear waveguides
  \cite{Kobayashi:17}. Both of these instances utilized CW lasers for
  pumps, and very narrowband signals (single photons, coherent states
  respectively) in their experiments.

  Temporal modes are field-orthogonal broadband wave-packet states of
  light occupying a certain frequency band \cite{tit66}. The CE of QFC
  devices can be made to be TM-selective, if the laser pump(s) have
  tailored pulse shapes \cite{Colin:12pra}. To see this, we express
  the equations of motion for QFC in terms of temporal-mode envelope
  functions $A_j(z,t)$, where the index $j$ labels the frequency band
  \cite{Colin:12pra,Reddy:2013ip}. For pulsed, three-wave mixing
  between pump ($p$-band), signal ($s$-band) and register ($r$-band)
  fields in a single-transverse-mode waveguide, the interaction
  Hamiltonian in a medium of length $L$ is
  \begin{equation}
    \hat{H}_I(t) = \gamma\int\limits_0^L dz A_p(t-\beta'_pz)\hat{A}_s(z,t)\hat{A}^\dagger_r(z,t) + h.a.\label{hint}
  \end{equation}
  \noindent where we assume energy conservation
  ($\omega_r=\omega_p+\omega_s$) and phase-matching at the band
  central/carrier frequencies ($\omega_j$). The parameters
  $\beta'_j\equiv\partial_\omega\beta(\omega)|_{\omega_j}$ are the
  group slownesses, or inverse-group velocities at said carrier
  frequencies. The temporal-mode envelope functions $\hat{A}_j(z,t)$ in
  the bands $j\in\{s,r\}$ have been elevated to photon creation
  operators
  \{$[\hat{A}_j(z,t),\hat{A}^\dagger_{j'}(z,t')]=\delta_{j,j'}\delta(t-t')$\}. Due
  to the linear nature of the field-evolution equations from the
  Hamiltonian in Eq. \ref{hint}, the operators can also stand for the
  classical TM-functions of weak-coherent pulses
  \cite{Titulaer:1966vb}. The pump is assumed to be a strong,
  nondepleting coherent state, and its pulse-mode envelope is assumed
  square normalized ($\int^\infty_{-\infty} dt|A_p(t)|^2=1$). The
  coupling strength $\gamma$ is proportional to the transverse-mode
  overlap integrals, the $\chi^{(2)}$-nonlinear coefficient, as well
  as the square-root of the pump pulse energy.

  Eq. \ref{hint} is identical in form to those governing a wide class
  of physical systems besides all-optical three-wave mixing. For
  instance, they can represent a pump-mediated interaction between an
  optical field and a collective Raman transition ``spin-wave'' in an
  atomic ensemble quantum memory \cite{Cirac:04ax,Nunn:07pra},
  implying that processes analogous to those studied here should also
  occur in such systems.

\begin{figure}[htb]
    \centering
    \includegraphics[width=\columnwidth]{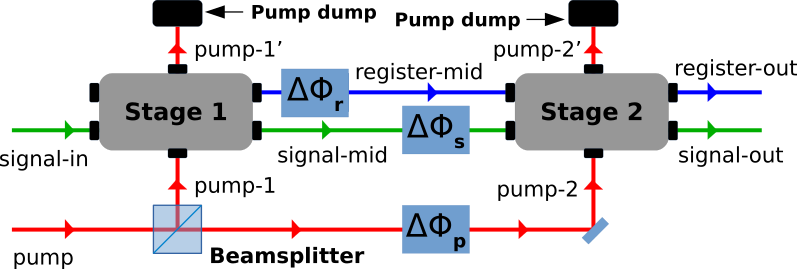}
    \caption{Schematic for a two-stage two-color optical Ramsey
      interferometer using three-wave mixing. If the medium
      dispersions are identical, then the pulse in the faster bands
      need to be delayed relative to those in the slower bands to
      ensure reinteraction in the second stage.}
    \label{fig01}
  \end{figure}

  The phenomenon can be expressed as a scattering matrix relating
  input-mode operators $\hat{A}_j(z=0,t')$ to output-mode operators
  $\hat{A}_j(z=L,t)$. It has been shown both theoretically and
  experimentally that for a wide range of system-parameter values
  ($\{\beta'_j,L,\gamma,\tau_p\}$, where $\tau_p$ is the pump pulse
  duration), at low effective-interaction strengths ($\tilde\gamma =
  \gamma\sqrt{L/|\beta'_r-\beta'_s|}$), the frequency conversion is
  TM-discriminatory, with the target TM being dependent on the complex
  shape of the pump-pulse envelope \cite{mej12b}. But increasing the
  interaction strength to reach higher CE imposes a trade-off
  condition between target-TM CE and TM discrimination
  \cite{Reddy:2013ip,Reddy:17}. This departure from high TM
  discrimination at large CEs is due to the effect of time-ordering
  corrections to the evolution \cite{Quesada:pra14}. The unitary
  evolution operator may be expressed in a Magnus expansion,
  \begin{equation}
    \hat{U}_I=\mathcal{T}\exp\left[-\frac{i}{\hbar}\int^\infty_{-\infty}dt\hat{H}_I(t)\right]=\exp(\hat\Omega_1+\hat\Omega_2+\hat\Omega_3+...),
  \end{equation}
  \noindent where $\mathcal{T}$ imposes time ordering. The first-order
  perturbative term $\exp(\hat\Omega_1)$ is, by definition, sans time
  ordering effects. It has been shown in theory by us and others
  \cite{Reddy:2014bt,Reddy:15pra,Quesada:16} that this limit to TM
  selectivity can be asymptotically overcome by preferentially
  weighing the first-order term in the expansion over the higher-order
  terms. This is achieved by coherently cascading multiple
  low-efficiency QFC stages in sequence. We have previously refered to
  this technique as temporal-mode interferometry (TMI). The basic
  schematic for a two-stage TMI is shown in fig. \ref{fig01}. The pump
  pulses for both stages are derived from a single pump beam to ensure
  constant relative phase. The net interferometric phase for the
  three-field process is $\Delta\Phi_p+\Delta\Phi_s-\Delta\Phi_r$.

  For this scheme to function, every stage needs to be highly
  TM-discriminatory, but not necessarily of high efficiency. The best
  parameter regime for high TM-discrimination at low CE is known to
  require that one of the weak bands (signal or register) copropagate
  with the pump pulse with a matching group velocity, and the other
  band's group velocity be drastically different
  \cite{silb11,Reddy:2013ip}. In our system, this condition is
  approximately satisfied through the use of an MgO-doped PPLN
  waveguide quasi-phasematched for Type-0 second-harmonic generation
  from $816.6$ nm to $408.3$ nm. We center our pump and signal bands
  at $821$ nm and $812.2$ nm respectively, and chose their bandwidths
  such that the pump and signal pulse durations ($\sim 500$ fs) are
  much larger than the pump-signal inter-pulse walkoff within the
  length of the medium, thus effectively mimicing the group-velocity
  matching condition \cite{Manurkar:16,Reddy:17}.

  The pump-signal walkoff scales linearly with medium length, implying
  that the group-velocity matching condition is better approximated in
  shorter waveguides. However, the single-stage TM-discrimination also
  requires that the register-band pulse walks off from the pump/signal
  pulse by a large amount \cite{Reddy:2013ip}. This necessitates that
  the ratio of the effective interaction time and the pump duration
  $\zeta = (|\beta'_r-\beta'_s|L/\tau_p)$ be much larger than unity
  \cite{Reddy:17}. We work with a $5$ mm long waveguide, yielding
  $\zeta\approx 10$, and a pump-signal walkoff of $< 40$ fs. Our pump
  and signal fields were derived from a homebuilt, Kerr-lens
  modelocked ultrafast Ti:sapphire laser producing pulses centered at
  $821$ nm and with a FWHM spectral-intensity bandwidth of $\sim 12$
  nm. The pulse repetition rate was $80$ MHz. The laser pulses were
  then directed into a 4f pulse shaper detailed in the methods
  section.

\begin{figure*}[tbh]
    \centering
    \includegraphics[width=\linewidth]{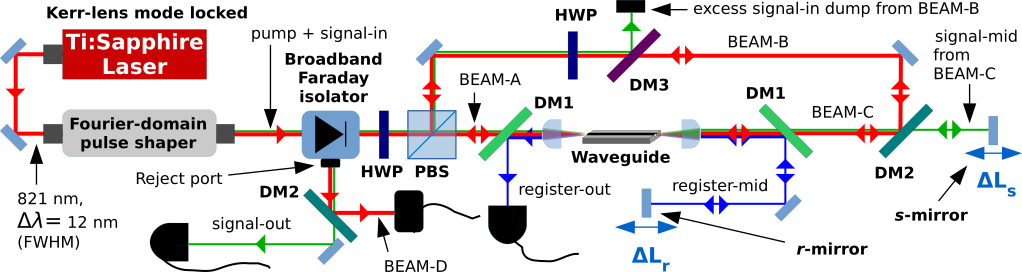}
    \caption{Two-stage, TM-selective optical Ramsey interferometer
      setup constructed by doublepassing all optical pulses through
      the same nonlinear MgO:PPLN waveguide. The pump pulse from the
      pulse shaper is split into two beams (A and B) at the polarizing
      beamsplitter (PBS). Pump-1 travels through the waveguide from
      left-to-right (beam-A to beam-C) and loops back to the PBS in
      the counterclockwise direction, and is eventually rejected into
      beam-D by the broadband Faraday isolator. Pump-2 travels along
      beam-B and goes through the waveguide from right-to-left (beam-C
      to beam-A), thus looping back to the PBS in the clockwise
      direction. The two pump pulses are within the waveguide at
      different times, forming the two stages. The signal-in pulse
      from beam-A copropagates with pump-1 through the waveguide, gets
      separated from the pump in beam-C through a dichroic mirror
      (DM2), and is then backreflected into the ``second-stage''
      waveguide from an end mirror labeled ``$s$-mirror.'' The
      signal-out pulse from beam-A is measured at the reject port of
      the Faraday isolator. The register-mid pulse generated from
      partial conversion of signal-in in the first-stage is
      backreflected into the second stage from the ``$r$-mirror.''
      All register-band pulses are seperated/combined with the pump-
      and signal-band pulses using dichroic mirrors (DM1). The
      ``$s$-mirror'' and ``$r$-mirror'' elements are mounted on
      high-precision linear translation stages for effective
      interferometric phase control via subwavelength displacements
      ($\Delta L_j=\lambda_j\Delta\Phi_j/2$ for $j\in\{s,r\}$). HWP
      stands for half-wave plate. DM3 is another dichroic
      element. Signal(register)-band beam paths are colored
      green(blue), while the pump-band beams are red.}
    \label{fig02}
  \end{figure*}

  The pump and signal-in pulses for the interferometer were generated
  by carving out their spectral components from the broadband
  ultrafast laser pulses in the pulse shaper \cite{Reddy:17}. We chose
  to work with modified Hermite-Gaussian functions with comparable
  bandwidths for the two-stage experiment. Specifically, we used three
  mode shapes defined as follows (with bandwidth parameter
  $\Delta\omega$, and normalization constants $N_j$):
\begin{align}
  HG0(\omega) &= \frac{1}{\sqrt{N_0}}\exp\left(-\frac{(\omega-\omega_0)^2}{2\Delta\omega^2}\right),\label{hg0eq}\\
  HG1(\omega) &= \frac{1}{\sqrt{N_1}}\left(\frac{\omega}{0.8\Delta\omega}\right)\exp\left(-\frac{(\omega-\omega_0)^2}{2(0.8\Delta\omega)^2}\right),\label{hg1eq}\\
  HG2(\omega) &= \frac{1}{\sqrt{N_2}}\left[2\left(\frac{\omega}{0.89\Delta\omega}\right)^2-1\right]\notag\\
  &\times\exp\left(-\frac{(\omega-\omega_0)^2}{2(0.8078\Delta\omega)^2}\right),\label{hg2eq}
\end{align}

\noindent where $\omega$ is the angular frequency. The width
modifications ensure mutual orthogonality whilst restricting total
bandwidth of all three modes to the same neighborhood ($\sim 2.5$
nm). Below, the pump and signal pulse shapes are denoted by $pj$ and
$sj$ respectively, where index $j\in\{0,1,2\}$ refers to shape $HGj$
from Eqs. \ref{hg0eq}-\ref{hg2eq}. The pump and signal-in pulses are
then sent towards the interferometer setup (see fig. \ref{fig02}).

The dispersion of the nonlinear media used in the two stages needs to
be identical, to ensure phasematching between the same central
wavelengths. We therefore reused the same waveguide twice, using a
back-reflection-based doublepass ``Michelson'' scheme detailed in the
figure. A broadband Faraday optical isolator enabled us to separate
the ``forward'' and ``backward'' propagating pulses from the two
stages, and measure the final outputs at its reject port. Longpass
dichroic mirrors (labeled DM2 in fig. \ref{fig02}) were used to
split/combine the $r$-band pulses from the $s$- and $p$-band
pulses. Similarly, other dichroic elements (DM1 and DM3 in
fig. \ref{fig02}) were employed to split/combine the different bands
in various permutations.

  Both the signal-mid and register-mid pulses were backreflected off
  of flat mirrors mounted on high-precision linear translation
  stages. These mirrors, refered to as ``$s$-mirror'' (for signal) and
  ``$r$-mirror'' (for register), were used to ensure both proper pulse
  overlap in the second stage, as well as to impart interferometric
  phase via subwavelength displacements. For example, a fine
  displacement of $\Delta L_s$ in the signal-mid arm would impart a
  phase of $\Delta\Phi_s = 2\Delta L_s/\lambda_s$ to the
  interferometer, with the factor of two resulting from the doublepass
  path-length change.

  All pulses were coupled into (and out of) the $5$ $\mu$m wide PPLN
  waveguide using $f = 11$ mm aspheric lenses. The coupling efficiency
  into the waveguide was around $30$\%. Since the conversion
  efficiencies in the two stages need to match each other, to ensure
  sufficient pump energy in the second stage, we derived the pump-2
  pulse afresh from the output of the shaper instead of backreflecting
  the transmitted pump-1 pulse. A half-wave plate and polarizing
  beamsplitter combination allowed us to independently redirect
  desired amounts of power into various beams. This freedom also
  enabled us to verify the identical operation of the waveguide in
  both directions by reproducing the single-stage results from
  \cite{Reddy:17} for both directions of propagation. A coupled
  average pump power of $0.47$ mW at a pulse rate of $80$ MHz yielded
  a CE of $0.5$ in both directions.

  The fringe visibility of the two-stage interferometer is sensitive
  to any imbalance in effective losses between the signal and register
  arms. Kobayashi et al. \cite{Kobayashi:17} have modeled all the loss
  channels in the two-color interferometer as disjoint unitary
  beamsplitters placed at various beam locations in the setup. We
  matched the losses (due to absorption in various elements, as well
  as inefficient coupling into waveguide) between the signal-mid and
  register-mid beams by inserting a spatially varying ND filter into
  the beam path of the better coupled (signal-mid) arm.

\begin{figure*}[bht]
    \centering
    \includegraphics[width=\linewidth]{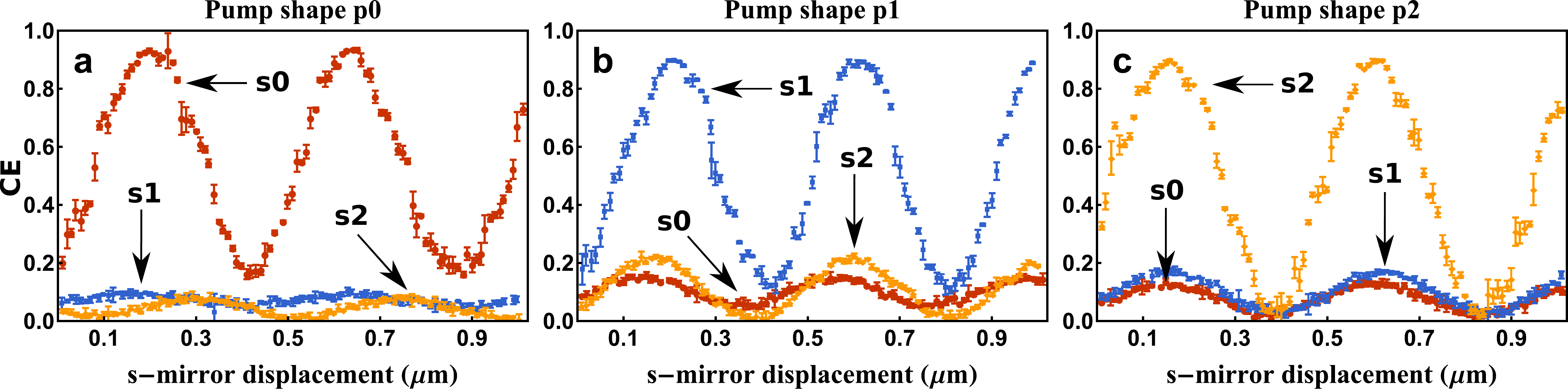}
    \caption{Two-stage, TM-selective optical Ramsey interferometric
      fringes visible in CE versus $s$-mirror displacement for
      pump-pulse shapes (a) p0, (b) p1, and (c) p2, with various
      signal TM shapes (s0, s1, s2). Shape functions defined in
      Eqs. \ref{hg0eq}-\ref{hg2eq}. The relative phase shifts amongst
      the plots are due to system drifts in between data runs.}
    \label{fig03}
  \end{figure*}

\begin{figure*}[tbh]
    \centering
    \includegraphics[width=\linewidth]{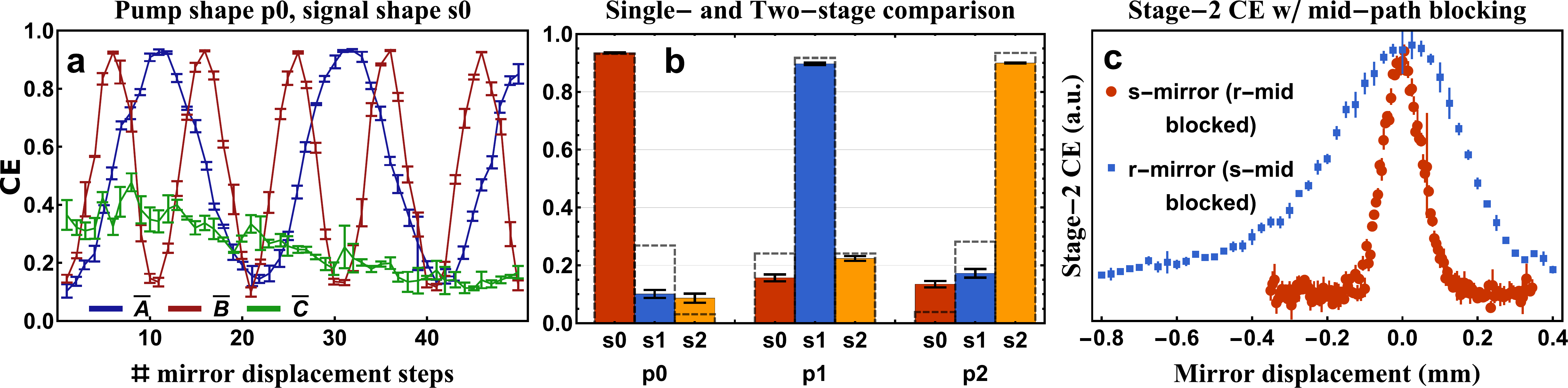}
    \caption{(a) Conversion efficiency versus
      combined-mirror-displacement for pump and signal shapes p0 and
      s0. $r$-mirror always moved in $+10$ nm steps. The legend items
      stand for $s$-mirror step sizes of $0$ nm ($\overline{\text
        A}$), $-10$ nm ($\overline{\text B}$), and $+10$ nm
      ($\overline{\text C}$). (b) Peak CE for various pump-signal
      shape combinations, and single-stage optimum levels for
      theoretically predicted exact Schmidt modes (dashed). (c) Cross
      correlation from Stage-2 only CE versus $s$($r$)-mirror
      displacement with register(signal)-mid beam blocked,
      demonstrating relative temporal widths. Pump and signal-in
      shapes were p0 and s0.}
    \label{fig04}
\end{figure*}

  Optical Ramsey-interference fringes were observed when recording the
  ``internal'' CE (defined as fraction of signal power depleted in the
  presence of pumps, which is independent of static losses) versus
  mirror displacements. Figure \ref{fig03} plots the CE versus
  $s$-mirror displacement for various combinations of pump and signal
  TM shapes. The spatial peak-to-peak period was found to be roughly
  half the signal-band wavelength, consistent with the backreflection
  setup. Correspondingly, $r$-mirror displacements resulted in a
  spatial period of half the register-band wavelength
  (fig. \ref{fig04}(a)).

  Figure \ref{fig04}(a) plots the CE fringe patterns for displacements
  of $r$-mirror alone ($10$ nm steps, legend item `$\overline{\text
    A}$'), both $r$-mirror and $s$-mirror in opposite directions ($10$
  nm and $20$ nm steps respectively, legend item `$\overline{\text
    B}$'), as well as both in the same direction (legend item
  `$\overline{\text C}$'). Here, direction is ``positive'' towards the
  waveguide, i.e. shortening path length. Combined mirror
  displacements in opposite directions halved the fringe period, while
  the CE was nearly unchanging for matching directional moves,
  confirming the relative signs in the net interferometric phase
  formula ($\Delta\Phi_p+\Delta\Phi_s-\Delta\Phi_r$). Note that the
  register frequency is not an exact harmonic of the signal frequency.

  The peak CE for the cases where the pump and signal TMs matched in
  shape far exceeded the shape-mismatched ones, as seen in
  fig. \ref{fig04}(b). Also plotted with dashed lines are the
  simulated single-stage CE for the theoretically predicted exact
  Schmidt modes at the pump powers required to match the maximum CE at
  each pump-pulse shape. The TM selectivity (roughly the contrast
  between targeted TM and other TMs) is enhanced for most shape
  combinations, even for signal TMs we used, which simply match the
  pump shapes (i.e. no attempt was made to optimize the signal pulses
  to match the exact Schmidt modes of the two-stage process). The
  single-stage results for the same waveguide and shaper slightly
  underperformed the numerical estimates \cite{Reddy:17}.

  The numerics assumed a parameter value $\zeta =
  (|\beta'_r-\beta'_s|L/\tau_p) = 10$. This must roughly equal the
  ratio of the temporal width of the register-mid pulse to that of the
  pump/signal-in/signal-mid pulse \cite{Reddy:2013ip}. We can estimate
  these widths by scanning the delay between pump-2 and the
  signal(register-mid) using $s$($r$)-mirror displacement and
  measuring the stage-2 CE while blocking the [register(signal)-mid]
  beam, as shown in fig. \ref{fig04}(c). The widths of pump-2 and
  signal-mid pulses will add in quadrature, since their group
  velocities are nearly equal. The pump-2 and idler-mid pulses,
  however, walk-off relative to each other in stage-2, resulting in a
  near-triangular CE curve. The data in fig. \ref{fig04}(c) indicates
  a $\zeta \sim 10$. It also demonstrates the expected temporal
  stretching of the register-mid pulse \cite{Reddy:2013ip}, related to
  the short SHG bandwidth of such waveguides \cite{Zheng:01,Zheng:02}.
  
\begin{figure}[bht]
    \centering
    \includegraphics[width=\linewidth]{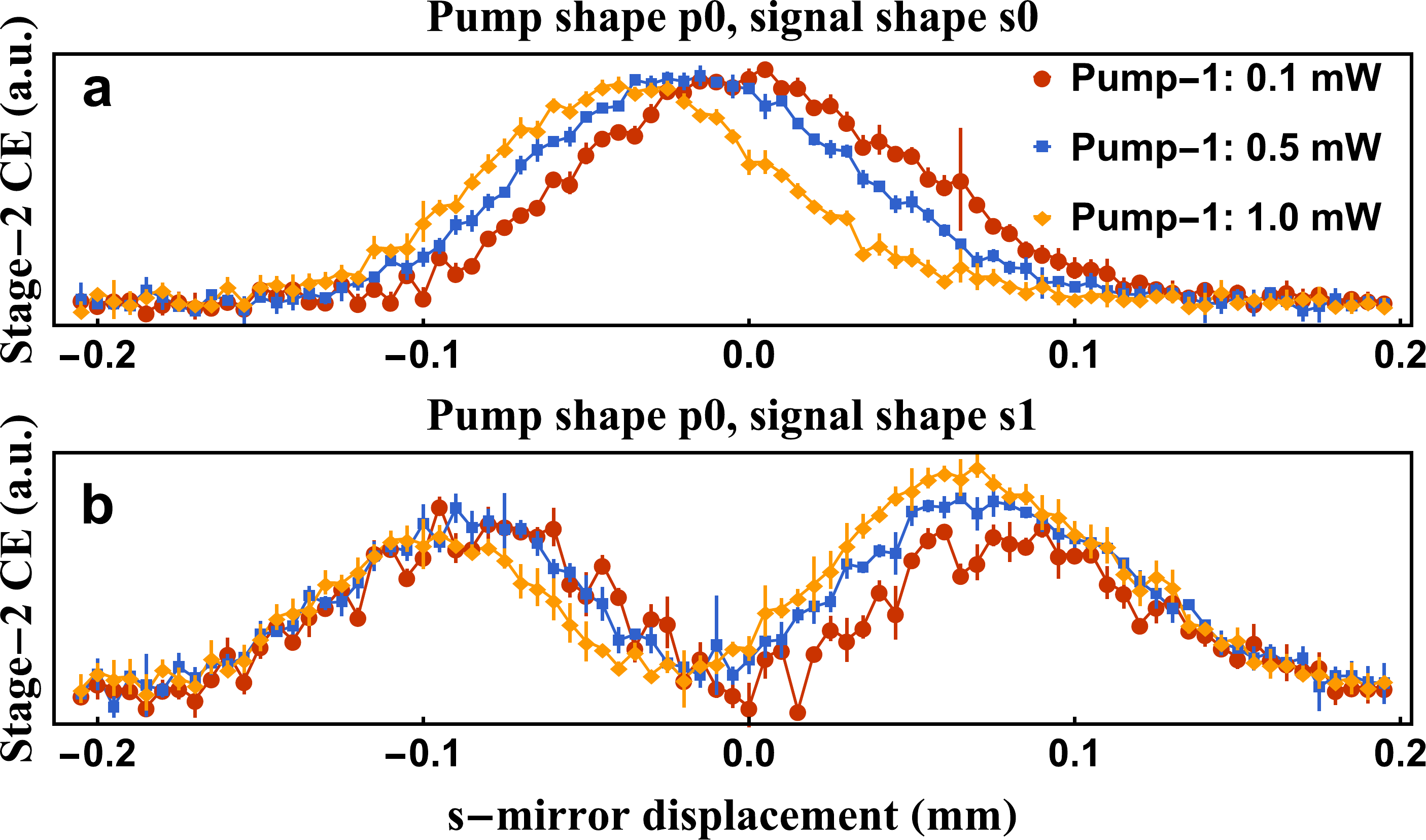}
    \caption{Stage-2 only CE versus $s$-mirror displacement with
      register-mid beam blocked for pump shape p0, and signal shape
      (a) s0, and (b) s1, at various pump-1 powers, and low pump-2
      power. The average pump-1 beam powers listed in the legend are
      for pulse rates of $80$ MHz. Positive displacement is towards
      the waveguide.}
    \label{fig05}
\end{figure}

  Stage-2 can separately be employed as a measurement device to
  demonstrate the effect of time ordering in the single-stage
  process. Figure \ref{fig05} plots the stage-2 only CE versus
  $s$-mirror displacement for signal shapes s0 and s1 with the
  register-mid beam blocked. Both pump-1 and pump-2 are shaped p0, and
  the pump-2 beam is weak (averaging $0.1$ mW at $80$ MHz pulse
  rate). Figure \ref{fig05}(a) shows that the signal-mid pulse
  temporally skews to earlier times, as well as compresses in width
  relative to signal-in pulse, as pump-1 power is increased. This
  illustrates the departure from the perturbative regime, and the
  distortion mimics the shape of the theoretically predicted
  single-stage output Schmidt modes \cite{Reddy:2013ip}. The same is
  true for signal shape s1 (fig. \ref{fig05}(b)), as the
  second-Schmidt mode for a Gaussian pulse skews to later times
  \cite{Reddy:2013ip}.

  The direction of the temporal skewing in fig. \ref{fig05}(a) is
  consistent with the sign of the signal-register relative group
  slownesses ($\beta'_r-\beta'_s > 0$). The register amplitude
  generated inside the waveguide lags behind the copropagating
  pump-signal pulses, causing an enhanced depletion of the later half
  of the signal pulse. The skewing direction is not due to the small
  difference in the group velocities of the pump and signal bands. It
  would remain invariant under an exchange of labels between the $p$-
  and $s$-bands (and a corresponding exchange of powers/amplitudes, as
  the ``pump''-band is defined as that which does not deplete).

  The two-stage optical Ramsey interferometer was not only able to
  enhance TM selectivity over the single-stage variants (even for
  unoptimized, pump-shape-matched signal TMs), but was able to achieve
  large CE with a more efficient pump-photon budget. The results in
  fig. \ref{fig03} and fig. \ref{fig04}(b) required beam powers of
  about $470$ mW coupled in for both pump-1 and pump-2 (at $80$ MHz
  pulse rates). Single-stage setups with the same waveguide could not
  reach such large CE for these net pump energies \cite{Reddy:17},
  even in theory \cite{Reddy:2013ip,Quesada:16}. This is due to the
  double passage through the nonlinear medium. The gain in interaction
  strength due to extension of medium length is superior to that from
  increased pump power as the latter has a lower time-ordering
  penalty. The likeness of the exact Schmidt modes to the pump-pulse
  shape at low CE
  \cite{eck11,silb11,Reddy:2013ip,Reddy:2014bt,Reddy:15pra,Reddy:17}
  enabled us to demonstrate enhancement in TM selectivtiy without
  having to optimize the signal-pulse shapes to match the said Schmidt
  modes.

  In conclusion, we have demonstrated temporal-mode-selective quantum
  frequency conversion in a two-stage optical Ramsey interferometer by
  utilizing shape-tailored strong laser pump pulses and weak coherent
  signal pulses, and double passing through a single nonlinear optical
  waveguide in a Michelson configuration. We showed that there is a
  significant selectivity enhancement over single-stage
  implementations in agreement with theoretical predictions. We note
  that by changing the phase of one of the fields inside the Ramsey
  interferometer, our two-stage device also acts as a
  frequency-conversion switch that operates on one selected temporal
  mode and (approximately) not on the others. We verified the temporal
  distortion of the signal pulse during nonlinear interaction
  predicted by theory, and related that to time ordering effects in
  general multi-field interaction systems that are governed by similar
  dynamical equations. This technique provides fruitful insight into
  the nature of time-ordering in pulsed scattering processes and paves
  the way towards designing a fully selective quantum pulse gate,
  thus, opening up the temporal mode basis space for quantum
  information processing.

  \section*{Methods}

  \noindent{\bf Experimental setup details.} The Fourier-domain, 4f
  pulse shaper used a $1800$ lines/mm holographic grating in
  near-Littrow configuration, and a cylindrical lens of focal length
  $250$ mm. We used a Meadowlark 8-bit, 2D, phase-only, reflective
  spatial light modulator of $1920\times 1152$ pixel resolution and
  array size of $17.6$ mm $\times 10.7$ mm in the Fourier plane of the
  pulse shaper. This allowed us to perform both amplitude and phase
  modulation using the Frumker-Silberberg first-order method
  \cite{Frumker:07,Reddy:17}.
  
  A Newport ISO-05-800-BB broadband Faraday optical isolator had a
  sufficiently flat transmission curve over a wide range of
  wavelengths for it to be deployed without imparting significant
  dispersion to the pump and signal-in pulses. The dichroic elements
  labeled DM1 in fig. \ref{fig02} are the DMLP650 longpass dichroic
  mirrors from Thorlabs. The transmission edges of Semrock FF01-810/10
  bandpass filters (DM2) and NF03-808E notch filters (DM3) were used
  to split/combine the $p$-band and $s$-band pulses from/with each
  other. The signal- and register- mirrors were mounted on Q-545.140
  closed-loop servo linear stages from Physik Instruments, which had a
  nomimal minimum step size of $6$ nm, and an encoder-based position
  read-out precision of $1$ nm. Low-profile flexture mounts from
  Siskiyou Inc. were used inbetween the two stages to construct a
  stable interferometer. Servo motors controlled by microcontrollers
  were used as beam blocks at various locations for convenience.
  
%

\section*{Acknowledgements}
This work was supported by a grant from the National Science
Foundation (NSF)(1521466), QIS - Quantum Information Science
Program. We owe much to the theoretical analysis of Dr. Colin
McKinstrie. We also thank Prof. Steven van Enk and Prof. Brian
J. Smith for discussions and suggestions regarding the experiment. We
thank Phil Battle and David Walsh AdvR for providing us with the
waveguide.

\section*{Author contributions}
DVR cenceived of the idea in the context of TM selectivity, and design
and built the whole setup. DVR also collected and analyzed the
data. MGR supervised the project closely. DVR prepared the manuscript,
with inputs from MGR.

\end{document}